\documentclass{llncs}

\usepackage{graphicx}
\usepackage{caption}
\usepackage{subcaption}
\usepackage{blindtext}
\usepackage[utf8]{inputenc}
\usepackage{amssymb,amsmath,array}
\usepackage{booktabs}
\usepackage[ruled,vlined,linesnumbered]{algorithm2e}
\usepackage[english]{babel}
\usepackage{listings}
\usepackage{xcolor}
\usepackage{blindtext}
\usepackage{times}
\usepackage{multirow}
\usepackage{longtable}
\usepackage{tabu}
\usepackage{balance}
\graphicspath{{media/}} 
\usepackage{tikz}
\usetikzlibrary{arrows, calc, fit, patterns, plotmarks, shapes.geometric,
shapes.misc, shapes.symbols, shapes.arrows, shapes.callouts, shapes.multipart,
shapes.gates.logic.US, shapes.gates.logic.IEC, er, automata, backgrounds,
chains, topaths, trees, petri, mindmap, matrix, calendar, folding, fadings,
through, positioning, scopes, decorations.fractals, decorations.shapes,
decorations.text, decorations.pathmorphing, decorations.pathreplacing,
decorations.footprints, decorations.markings, shadows}

\usepackage[
bookmarks=false,
breaklinks=true,
colorlinks=true,
linkcolor=black,
citecolor=black,
urlcolor=black,
pdfpagelayout=SinglePage,
pdfstartview=Fit
]
{hyperref}                                           
\hypersetup{
    pdftitle={Distributed Holistic Clustering on Linked Data},
    pdfauthor={Markus Nentwig, Anika Groß, Maximilian Möller, Erhard Rahm},
    % ...
}

\usepackage[capitalise,nameinlink]{cleveref}
\crefname{section}{Section}{Section}
\crefname{table}{Table}{Table}
\Crefname{section}{Section}{Sections}
\crefname{figure}{Fig.}{Fig.}
\Crefname{figure}{Figure}{Figures}

\newcommand{\eg}{e.\,g.}

\DeclareRobustCommand{\C}{\mathcal{C}}

\newcommand{\E}{\mathcal{E}}

\newcommand{\V}{\mathcal{V}}
\newcommand{\G}{\mathcal{G}}

\begin{document}
\title{Distributed Holistic Clustering on Linked Data}
                                                                                
\author{Markus Nentwig\inst{1}\inst{2} \and Anika Groß\inst{1,2} \and Maximilian
Möller\inst{1,3} \and Erhard Rahm\inst{1,2}
}
\authorrunning{Markus Nentwig et al.}

\institute{Database Group, University of Leipzig
    \and
    \email{nentwig|gross|rahm@informatik.uni-leipzig.de}
    \and
    \email{m.moeller@studserv.uni-leipzig.de}
}

\maketitle

\begin{abstract} 

Link discovery is an active field of research to support data integration in the
Web of Data. Due to the huge size and number of available data sources,
efficient and effective link discovery is a very challenging task. Common
pairwise link discovery approaches do not scale to many sources with very large
entity sets. 
We here propose a distributed holistic approach to link many data sources based
on a clustering of entities that represent the same real-world object. Our
clustering approach provides a compact and fused representation of entities, and can
identify errors in existing links as well as many new links. We support a
distributed execution of the clustering approach to achieve faster execution
times and scalability for large real-world data sets. We provide a novel gold
standard for multi-source clustering, and evaluate our methods with respect to
effectiveness and efficiency for large data sets from the geographic and
music domains. 
\end{abstract}

\section{Introduction}
\label{sec:introduction}

Linking entities from various sources and domains is one of the crucial steps to
support data integration in the Web of Data. A manual generation of links is
very time-consuming and nearly infeasible for the large number of existing
entities and data sources. As a consequence, there has been much research effort
to develop link discovery (LD) frameworks~\cite{Nentwig2017} for automatic link
generation. Platforms like \texttt{datahub.io} and \texttt{sameas.org} or
repositories such as LinkLion~\cite{Nentwig2014} and BioPortal~\cite{Noy2009}
collect and provide large sets of links between numerous different knowledge sources. They are valuable resources to improve the availability and re-usability of links in applications and avoid an expensive re-determination of the links. However, existing inter-source mappings can be incomplete and so far many sources are not interrelated at all.

Due to the huge number and size of available knowledge sources, link discovery is still a very challenging task. 
It is particularly complex to ensure high link quality, i.e., the generation of
correct and complete link sets. Existing link repositories cover only a small
number of inter-source mappings and automatically generated links can be
erroneous in many cases~\cite{Faria2014a,Ngomo2011}. Despite the huge number of
sources to be linked, most LD tools focus on a pairwise (binary) linking of
sources. However, link discovery approaches need to scale for n-ary linking
tasks where more than two sources need to be matched, as well as for an
increasing number of entities and sources that are added to the Web of Data over
time~\cite{Rahm2016}. 


To address these shortcomings we recently proposed an approach to cluster linked
data entities from multiple data sources into a holistic representation with
unified properties~\cite{Nentwig2016}. The method combines entities that refer
to the same real world object in one cluster instead of maintaining a high
number of binary links for $k$ sources. The approach is based on existing
\texttt{owl:sameAs} links and can deal with entities of different semantic types
as they occur in many sources (\eg, geographical datasets
contain various kinds of entities such as countries, cities, lakes). It can
further eliminate wrong links from existing link sets and identifies many new
links, \eg, for previously unconnected sources. The clustered representation of entities with their properties and a cluster representative is more compact and comprehensive, allows central maintenance and access, and facilitates the integration of new data sources and entities. The clustering-based approach complements existing link discovery approaches to provide a compact and fused representation of entities. Initial clusters can further serve as a basis for an incremental extension of clusters by newly added entities as outlined in~\cite{Rahm2016}.  

Considering the huge size and number of sources to be linked, scalability
becomes a major issue. Workflows for linking and/or clustering of entities
usually comprise several complex phases such as similarity computation to find
similar entities or clusters that need to be merged. Therefore, a distributed
realization of holistic clustering approaches for $k$ data sources would be
valuable to improve its efficiency. In particular, complex operations should be
executed in parallel in a distributed environment in order to reduce execution
time and to enable scalability for large real-world datasets. Big Data
frameworks like Apache Spark~\cite{Zaharia2012} or Apache
Flink~\cite{Carbone2015} provide fast execution engines to process very large
datasets in a distributed environment. These generic frameworks allow to define
complex data processing workflows using common operators like $map$ or $join$ as
well as user-defined functions. 
So far there has been only few related work considering distributed clustering for linked data and  data integration workflows. 
With regard to the ever increasing amount of data that needs to be linked and integrated in typical big data processing
workflows, it is essential to develop scalable solutions for link discovery and holistic entity
clustering.



A further and long-standing problem is the poor availability of reference datasets that can be used as gold standard for quality evaluations. Since LD tools usually focus on pairwise matching, the few existing benchmarks 
cover 
links between two data sources. However, there is an increasing need to evaluate
the effectiveness of multi-source clustering, as holistic data integration
scenarios like the construction of knowledge graphs gains increasing research interest~\cite{dong2014vault,nickel2015review}. The creation of new reference datasets for quality evaluation of n-ary linking and clustering approaches can be useful for the community and support the development of improved holistic clustering approaches.


In this work, we study distributed holistic clustering as well as its evaluation on real-world data from different domains. We make the following major contributions:
\begin{itemize}
    \item We present a distributed holistic clustering approach for linked data to enable an effective and efficient clustering of large entity sets from many data sources. The implementation is based on the distributed 
		data processing system Apache Flink. 
		\item We provide a novel reference dataset for multi-source clustering from the geographic domain. 
		We evaluate the effectiveness of our approach with the new gold standard and a further artificial dataset from the music domain.
    \item We further evaluate the efficiency and scalability of the distributed holistic clustering for very large datasets with millions of entities from the two domains.  
\end{itemize}

The remainder of the paper is organized as follows. We first introduce
preliminaries w.r.t. link discovery and the holistic
clustering approach (\cref{sec:prelim}). In \cref{sec:impl} we present the 
implementation of the distributed holistic clustering. We present the
clustering gold standard in Section \ref{sec:benchmark} and show
evaluation results in \cref{sec:evaluation}. Finally,
we discuss related work in \cref{sec:relwork} and
conclude in \cref{sec:conclusion}.

\section{Preliminaries} 
\label{sec:prelim}

\subsection{Link Discovery}

Link discovery (LD) has been studied intensively and there is a large number of approaches and prototypes available~\cite{Nentwig2017}. 
LD tools usually identify \texttt{owl:sameAs} links between entities that denote the same real-world object. 
However, link discovery might also include the identification of links with other semantics such as the association \texttt{dbp:birthPlace} between a person and a city. 
Entities are described by a unique URI, \eg,  
\texttt{http://dbpedia.org/page/Leipzig}, and can have different semantic types,
\eg,
\texttt{dbo:Country} or \texttt{dbo:City} for the geographic domain.
Each entity can be further described by specific properties from ontologies
such as \texttt{rdfs:label}. 
The matching process usually covers the computation of similarity values for
pairs of entities, and candidate links are subsequently filtered by selection
strategies based on the link similarity. 
The herein presented approach complements existing binary LD approaches for
linking two data sources by allowing for a distributed holistic clustering of
entities from $k$ different data sources $S$. 
Same-as links created via binary LD constitute a mapping
$M_{i,j}=\{(e_1,e_2,sim)|e_1\in S_i, e_2\in S_j, sim[0,1],i \neq j, 1\le i,j \le
k\}$. We reuse existing mappings between different data sources
$\mathcal{M}=\bigcup_{i,j=1}^{k}{M_{i,j}}$ as input for the holistic
clustering.


\subsection{Holistic Clustering Workflow}

\begin{figure}[t!bp]
	\centering
		\includegraphics[width=1.00\textwidth]{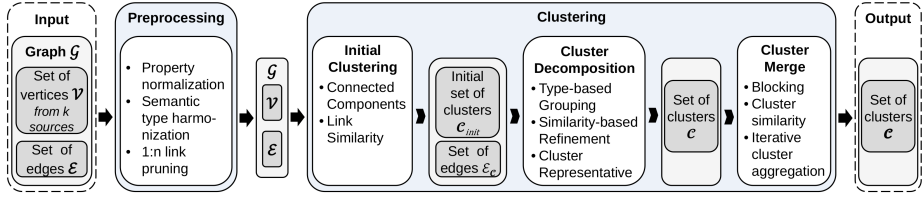}
	\caption{Holistic Clustering Workflow}
	\label{fig:workflow}
	\vspace{-2mm}
\end{figure}

In our previous work~\cite{Nentwig2016} we presented the holistic clustering
workflow shown in Figure~\ref{fig:workflow}. Here we present the realization of
the holistic clustering in a distributed environment for improved performance
and scalability of the approach. We therefore use the distributed processing
system Apache Flink as well as its graph-processing API Gelly. Our
implementation is based on the property graph model. The input of the workflow
is a graph $\G = (\V,\E)$ with a set of vertices (entities) $\V = \{v\}$ with
each vertex having an unique identifier and domain-specific properties, \eg, $v
= \{$id, label, source, type, \dots$\}$, as well as a set of edges (links) $\E =
\{e\}$ where each edge has a source and a target vertex id in $\V$. Edges may
also have additional properties like a similarity value.
The input graph $\G$ is a similarity graph constructed based on an existing
mapping of input links $\mathcal{M}$, \eg, from an publicly available link repository.
For our representation of clusters we define a set of clusters $\C = \{C\}$
where each $C \in \C$ has a cluster id $c_{\text{id}}$, a set of contained
vertex ids from $\V$ and a cluster representative $r$ containing unified
properties like a label, a semantic type and the original data sources, \eg, $r
= \{\text{label},\{v_1,v_2,\dots\}, \{S_i, S_j, \dots\}, \{t_1, t_2, \dots\},
\dots\}$.  Although input links might be associated with similarity values, we
compute new similarities between vertices within clusters (intra-cluster edges
$\E_\C$) to enable the use of sophisticated measures and to ensure comparability
of link similarities within our workflow. 

The basic holistic entity clustering~\cite{Nentwig2016} employs a four-phase
approach (see Figure~\ref{fig:workflow}). During \emph{Preprocessing} we
normalize properties for vertices and harmonize their semantic types. The
harmonization is necessary since semantic type information can differ
substantially between different sources, and might also be missing. We therefore
use background knowledge about type equivalences and comparability to harmonize
original types to more general ones, \eg, \textit{city}, \textit{town} and
\textit{suburb} refer to the more general type \textit{settlement}. For
duplicate-free data sources there should be at most one equivalent entity in
another data source. For entities with several links to entities from the same
source (1:n links), we therefore only keep the best link for the considered
entity in order to ensure the one-to-one cardinality.  As an \emph{Initial
Clustering} we compute connected components for the input graph $G$. Within each
initial cluster $C \in \mathcal{C}$ we then compare all covered vertices based
on different similarity measures to obtain a set of intra-cluster edges $\E_C$
and similarity values for the subsequent steps. In the \emph{Cluster
Decomposition} phase we refine the clusters by splitting them according to
semantic type information and edge similarities within the cluster. Each cluster
should only contain vertices of the same semantic type, and edges within a
cluster have to exceed a certain similarity threshold.  Vertices that do not
fulfill these criteria constitute a new cluster. For each cluster we create a
cluster representative by combining property and data source information from
the covered vertices. The final \emph{Cluster Merge} phase aims at combining
smaller but similar clusters. Therefore, we first compute similarities between
different clusters, and then iteratively merge the best matching cluster
combinations, respectively, checking constraints like the maximal possible
cluster size $k$ and already covered data sources within a cluster. In the here
presented implementation we further employ a new blocking step prior to the
matching of the cluster representatives in order to reduce the search space and
avoid unnecessary comparisons. 

In contrast to~\cite{Nentwig2016}, this
paper outlines a distributed implementation of the holistic clustering approach
as well as a comprehensive evaluation of its quality and efficiency.

%


\section{Implementation of Distributed Holistic Clustering}
\label{sec:impl}

In this section we outline the workflow and implementation for our distributed
holistic clustering approach based on the big data stream and batch processing
system Apache Flink~\cite{Carbone2015}. We first introduce the used
transformation operations from Apache Flink as well as graph datasets provided
by Flink's graph processing engine Gelly (Section \ref{sec:impl:flink}).  We
then present the transformation and adaptation of the holistic clustering
approach towards a distributed processing workflow using Apache Flink/Gelly
(Section \ref{sec:impl:holistic}). 

\subsection{Apache Flink and Gelly API} 
\label{sec:impl:flink}
For our approach we  make use of the batch processing part of Apache Flink. 
It provides the DataSet API and well-known dataset transformations 
like \emph{filter}, \emph{join}, \emph{union}, \emph{group-by} or
\emph{aggregations} (relational databases) and \emph{map}, \emph{flat-map} and
\emph{reduce} (MapReduce paradigm). Special in-memory, distributed data
structures called DataSets store data within Flink programs. DataSets can be 
manipulated based on so called transformations that return a new DataSet. 
Some transformation operations make use of user-defined functions (UDFs) 
and allow for customized definitions  
how DataSet values need to be changed. 
Flink optimizes the execution of succeeding transformations, employs lazy evaluation
and avoids intermediate materialization of results.
A set of libraries provides additional functionality
for Flink such as complex event processing, graph processing or machine
learning. We will make use of the graph processing engine (Gelly) in our holistic 
clustering workflow. In particular, we employ Gelly graphs containing a DataSet
of vertices and edges

{\small
\begin{verbatim}
class Graph<K, VV, EV> graph
    DataSet<Vertex<K, VV>> vertices;
    DataSet<Edge<K, EV>> edges;
}
\end{verbatim}
}


The complex data types  \texttt{Vertex} and \texttt{Edge} are inherited from
the Flink \texttt{Tuple} classes \texttt{Tuple2<K,VV>} (type \texttt{K} as vertex id,
\texttt{VV} as vertex value), \texttt{Tuple3<K,K,EV>} (source vertex id,
target vertex id (each type \texttt{K}) and \texttt{EV} as edge value),
respectively.  Operators like \emph{join}, \emph{filter} or \emph{group-by}
rely on tuple positions (starting from 0), \eg,

{\small
\begin{verbatim}
vertices.join(edges)
    .where(0).equalTo(1)
    .with((vertex, edge) -> new Tuple1<>(edge.getSimilarity))
    .filter(tuple -> tuple.f0 >= 0.9);
\end{verbatim}
}

will join all edges with the vertices where the vertex id (position 0 in
\texttt{vertices}) equals the target id of the edge (position 1 in
\texttt{edges}) and returns the similarity value if the accompanied filter
function is evaluated and returns true. 
Please see \cref{tab:ops} for more
detailed explanations on transformations that we use within our approach. 
Depending on the physical data distribution, data needs to be shuffled across
cluster nodes to execute transformations such as join or group-reduce requiring
a reorganisation of data w.r.t. the used (grouping) key. Strategies like
repartition/map-side join or group-reduce/combine can be applied to reduce the
necessary network traffic. 

Besides the used graph data model we benefit from Flink's and Gelly's abstract
graph processing operators like graph neighborhood aggregations or abstracted
models for iterative computations. In particular, we will make use of the Flink
delta iteration in different variations, \eg, vertex-centric iteration
(Pregel~\cite{Malewicz2010}), gather-sum-apply computation
(PowerGraph~\cite{Gonzalez2012}) or custom implementations for delta iterations
 as discussed in the following sections.

{\small
\begin{table*}[t]
    \centering
    \begin{tabular}{lm{10cm}}
        \toprule
        Operator & Description\\
        \midrule
        (Flat)Map & Map- and FlatMap transformations apply a UDF on each element
        of the DataSet. Map functions emit exactly one resulting element per
        input while FlatMap functions may emit arbitrary result elements
        (including none). \\ 
        & \texttt{input.map(udf: IN -> OUT)}\\
        \midrule
        Filter & Return all DataSet elements for which the UDF returns \texttt{true}. \\
        & \texttt{input.filter(udf: IN -> Boolean)}\\
        \midrule
        (Flat)Join & A Join transformation combines elements of two DataSets
        with equal values on (tuple) key(s) and creates exactly one result
        element for Join and arbitrary
        result elements for the FlatJoin based on a UDF .\\
        & \texttt{leftInput.join(rightInput).where(leftKeys)} \\
        & \texttt{.equalTo(rightKeys).with(udf: (left,right) -> OUT)}\\
        \midrule
        ReduceGroup & Elements in a DataSet can be grouped on keys of the
        DataSet fields. Each group executes a user-defined group-reduce function.\\
        & \texttt{input.groupBy(keys).reduceGroup(udf: IN -> OUT)} \\
        \midrule
        Aggregate & Aggregations like \texttt{sum}, \texttt{min}, 
        \texttt{max} or UDFs can be applied to grouped DataSets \\
        & \texttt{input.aggregate(SUM, key);} \\
        \bottomrule
    \end{tabular}
    \caption{Apache Flink DataSet transformations used for holistic clustering
        approach.}
    \label{tab:ops}
\end{table*}
}

\subsection{Holistic Clustering}
\label{sec:impl:holistic}


In this section, we will discuss the transformation and adaptation 
of the holistic clustering workflow towards a
distributed processing workflow in Apache Flink. 
From a high-level perspective, we read input vertices and edges 
into a Gelly graph and apply a set of transformation operators to
generate entity clusters. Intermediate results are represented as Gelly
vertices, edges and graphs and can be written to disk, \eg, as JSON. Within
complex transformations we remove unneeded properties using Flink \texttt{TupleX}
representation instead of \texttt{Vertex} or \texttt{Edge} to reduce the amount
of network traffic and memory consumption. 
We illustrate the 
workflow steps using the running example in \cref{fig:example}. There are six input edges $\E$ and seven 
input vertices $\V$ further described by a label ($l_1$, $l_2$, \dots), the originating data source from $S$ 
and colored dependent on their semantic type ($t1$, $t2$ or no type). 

\subsubsection{Preprocessing} 

\begin{figure}[t!bp]
	\centering
		\includegraphics[width=1.00\textwidth]{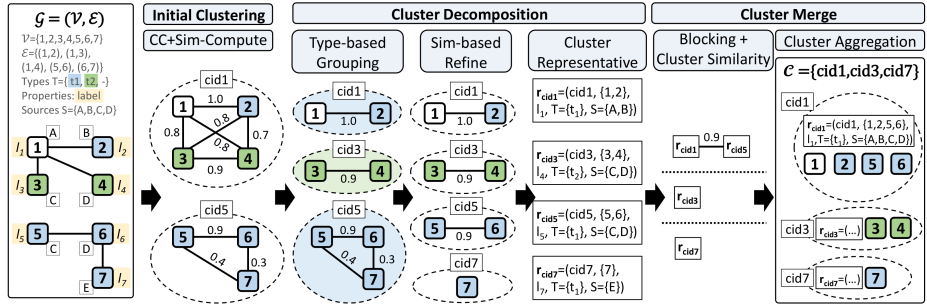}
	\caption{Example clustering workflow. }
	\label{fig:example}
\end{figure}

During preprocessing we apply several user-defined functions on
the input graph, \eg, to harmonize semantic type information, remove
inconsistent edges and vertices and normalize the label property value.
First, we compute similarities only for given input edges based on vertex property values. 
For each vertex, we carry out a consistency validation using grouping on
adjacent vertices and associated edges, and further remove neighbors 
with equal data sources (for details see \cite{Nentwig2016}).
For vertices with multiple semantic type values we ensure that grouping on the
type supports overlapping type information for later processing.
 For clarity reasons, we do not use multiple types in our running example. 
We also omit the preprocessing step in the example (\cref{fig:example}) 
and directly start with the preprocessed input graph $\G = (\V,\E)$. 

\subsubsection{Initial Clustering}

To determine initial clusters, we use a gather-sum-apply operator to determine 
the connected components (CC) within $\G$ and assign a cluster id to each vertex. 
In the example, vertices 1-4 obtain obtain cluster ids \emph{cid1} and
vertices 5-7 \emph{cid5}. With these cluster ids, we generate all missing 
links within each cluster. We therefore use a CoGroup function on cluster ids 
of all vertices, combine vertices within clusters to get intra-cluster edges $\E_\C$ 
and compute similarities for these edges. 
We apply a combination of similarity measures on different properties such as a 
linguistic similarity on labels and other properties and possibly further 
domain-specific measures like a normalized geographical distance.

\subsubsection{Type-based Grouping}

\begin{figure*}[t!bp]
    \centering
        \includegraphics[width=1.0\textwidth]{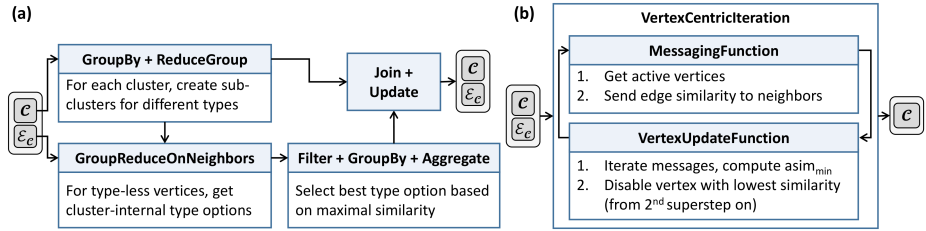}
    \caption{Sub-workflows with operators for type-based grouping (a) and
        similarity-based refinement (b).}
    \label{fig:typegroup-simsort}
\end{figure*}

Type-based grouping is the first part of the decomposition phase 
to split clusters into sub-components dependent on the compatibility of semantic types. 
\cref{fig:typegroup-simsort}\,a
shows the sequence of applied transformations and short descriptions. 
First the initially clustered vertices are grouped on
their cluster ids. Each cluster group executes a ReduceGroup function to assign
new cluster ids based on different semantic type, \eg, in the example vertex 3 and 4 are
separated from vertex 2. All vertices without type (like vertex 1) require a special
handling. We apply GroupReduceOnNeighbors (a Gelly CoGroup function to
handle neighboring vertices and edges for each vertex) to produce
\texttt{Tuple4} objects for vertices with missing semantic type. 
In our example, vertex 1 creates a \texttt{Tuple4<>(id,sim,type,cid)} 
for each of its outgoing edges ($(1,2),(1,3),(1,4)$), 
namely \texttt{Tuple4<>(1, 1.0, t\_1,cid1)} for edge $(1,2)$ and 
\texttt{Tuple4<>(1, 0.8, t\_2, cid2)} for edge
$(1,3)$ and $(1,4)$.  Now, grouping on the vertex id executes an
aggregation function for each group to return the tuple with the highest
similarity per vertex, which is \texttt{Tuple4<>(1, 1.0, t\_1, cid1)} for vertex 1 in our running
example.  
Finally, the vertices from the initial group-by (see \cref{fig:typegroup-simsort}\,a are
joined with the remaining \texttt{Tuple4} objects to update vertices without
semantic type with their new cluster id, \eg, vertex 1 is assigned to cluster
$cid1$. The result of the type-based grouping is a set of clusters with intra-cluster edges. 

\subsubsection{Similarity-based Refinement}


We further decompose clusters by removing non-similar entities that do not match
with other entities in their cluster. This part is realized using the Gelly
vertex-centric iteration. Basically, the implemented vertex-centric iteration
switches between a custom MessagingFunction and a VertexUpdateFunction (see
\cref{fig:typegroup-simsort}\,b for details), an iteration round is called
superstep.  In the first round, all vertices are active and therefore send
messages to all their neighbors. Messages are \texttt{Tuple3} objects containing
the originating id, the edge similarity and an average edge similarity $asim$
over all incoming messages (which is $0$ in the first iteration). 
Starting with the second iteration, we illustrate the sent messages for vertex 7
in cluster $cid5$ in our example: vertex 5 sends \texttt{Tuple3<>(5, 0.4, 0.65)}
to 7, vertex 6 sends \texttt{Tuple3<>(6, 0.3, 0.6)} to 7 and vertex 7 sends
messages to 5 and 6, resulting in the average similarity $asim = (0.4 + 0.3) / 2
= 0.35$ for vertex 7.  Now in each cluster the vertex with the lowest $asim$
will be deactivated (and is therefore excluded from the cluster) given that this
$asim$ is below a certain similarity threshold. In our example in
\cref{fig:example}, vertex 7 with $asim = 0.35$ will be deactivated and isolated
into cluster $cid7$.  The vertex-centric iteration terminates based on these 
criteria: First, vertices only send messages when their own vertex value was
updated in the current round. Second,  all deactivated vertices will never send
messages again. Lastly, a maximum number of iteration rounds can be set.

\subsubsection{Cluster Representative}

As final step in the decomposition phase, we create a representative for each cluster. 
Therefore, a GroupReduce function based on the cluster id is used to combine property
values to a unified representation. To determine the cluster representative, we aggregate cluster 
information about the covered data sources, 
semantic types and contained vertices and select the best values for certain properties such as label or 
geographic coordinates using the GroupReduce on clusters. 
Detailed examples for representants are shown in \cref{fig:example} ($r_{cid1},
r_{cid3}, r_{cid5}$ and $r_{cid7}$).


\subsubsection{Cluster Merge}

\begin{figure*}[t!]
    \centering
        \includegraphics[width=1.0\textwidth]{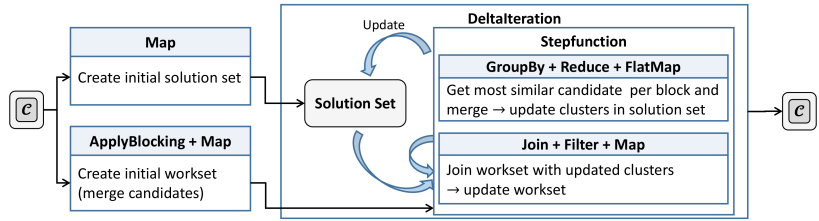}
    \caption{Sequence of transformations for the cluster merge using Flink DeltaIteration}
    \label{fig:merge}
\end{figure*}

For the final merge phase in our distributed holistic clustering workflow 
we use the Flink DeltaIterate operator together with user-defined functions. 
The main operators are sketched in \cref{fig:merge}. During merge, we iteratively 
aggregate highly similar (likely small) clusters into larger ones. 
With the creation of representatives for each cluster, we already reduced the
amount of entities to handle in the merge step. Due to the fact that we
potentially compare every cluster with every other cluster, the quadratic
complexity can become a problem for very large cluster sets. Thus we employ
blocking strategies and avoid unnecessary comparisons. Currently we implement
standard blocking on specified property values such as using the first letters
of the label as blocking key.  We avoid further unnecessary comparisons since we
do not compare representatives with incompatible types based on semantic type
and check for already covered data sources since we assume duplicate-free data
sources.  In our example in \cref{fig:example} three blocks are shown for the
blocking: $r_{cid1}$ and $r_{cid5}$ need to be compared, such that a triplet
$(r_{cid1}, 0.9, r_{cid5})$ is created as a merge candidate. For $r_{cid3}$, no
merge candidate is created because there is no other representative with type
$t_2$, whereas $r_{cid7}$ is in a separate block due to its dissimilar label
compared to other representatives. 

The delta iteration is started with an initial solution set 
containing the previously determined clusters
and an initial workset (merge candidates) as seen in \cref{fig:merge}. Within
each iteration of DeltaIterate, a custom step function works on the current
workset to update the workset and to generate changes on the current solution
set. The next workset and the updates to the solution set are then passed to the
next iteration step. In detail, the workset is grouped by the
blocking key and for each block the triplet with the highest similarity
exceeding the minimum similarity is selected using a custom Reduce function. For
our running example, $(r_{cid1}, 0.9, r_{cid5})$ is the best merge candidate and
therefore merged with a custom FlatMap function. The new cluster $r_{cid1}$
contains combined values for properties like sources $S=\{A,B,C,D\}$, the list
of contained vertices ($\{1,2,5,6\}$) and unified properties like the label
$l_1$. This will directly affect the cluster representatives in the solution set,
and the already merged cluster $r_{cid5}$ will be deactivated in the solution set.
Now the merge candidates within the workset are adapted based on 
changed clusters within the iteration step (see \cref{fig:merge} solution
set). Deactivated cluster representatives are replaced by the
appropriate new cluster (here $r_{cid1}$), and merged triplets ($(r_{cid1}, 0.9,
r_{cid5})$) as well as generated duplicate triplets are removed from the
workset. Again, triplets are discarded if the data sources for the participating
clusters overlap or exceed the maximum possible number of covered sources.
The delta iteration ends either when 
the workset is empty (default, and true for our running example after the first
iteration) or a maximum number of iterations took place. 
Note, that for larger datasets, parts of the dataset will converge faster to a solution,
when clusters can not be merged anymore. These parts will not be recomputed in
following iterations, such that only smaller parts of the data will be handled. 


\section{Evaluation}
\label{sec:evaluation}

In the following we evaluate our distributed holistic clustering approach
w.r.t. effectiveness and efficiency for datasets from the 
geographic and music domains. We first describe details of the used datasets (Section~\ref{sec:data}). 
We then present a novel reference dataset for multi-source clustering and describe its creation process (Section~\ref{sec:benchmark}). 
Based on this real-world and a further artificial reference dataset, we evaluate the effectiveness 
of our approach (Section~\ref{sec:results:qual}). Based on large datasets from
the geographic and music domains we also analyze the efficiency and scalability of the distributed holistic clustering approach.

\begin{table*}[b]
    \vspace{-5mm}
    \begin{center}
		\setlength{\tabcolsep}{1mm}
    \begin{longtabu}{lccccccr}
    \rowfont{\bfseries}
        \toprule
        domain                      & entity properties       &dataset & \#entities      &\#sources & \#correct links & \#clusters\\
				\hline
        \multirow{2}{*}{geography}  & label, semantic type,       &DS1     &$3{,}054$        &$4$       & $4{,}391$	   & $820$\\
                                    & longitude, latitude  &DS2     & $1{,}537{,}243$ &$5$       & -             & - \\
				\hline
        \multirow{3}{*}{music}      & artist, title, album,  &DS3     & $19{,}375$       &$5$       & $16{,}250$    & $10{,}000$ \\
                                    & year, length,          &DS4     & $1{,}937{,}500$  &$5$       & $1{,}624{,}503$& $1{,}000{,}000$ \\
                                    & language, number       &DS5     & $19{,}375{,}000$ &$5$       & $16{,}242{,}849$& $10{,}000{,}000$ \\
        \bottomrule
    \end{longtabu}
    \end{center}
    \caption{Overview of evaluation datasets. Number of resulting clusters and
    deduced correct links are given for reference datasets.}
    \label{tab:datasets}
    \vspace{-5mm}
\end{table*}

\subsection{Datasets}
\label{sec:data}

To evaluate the distributed holistic clustering, we use five datasets of
different sizes from the music and geographic domains. \cref{tab:datasets} shows
the available property values for entities, the number of covered entities and
sources, as well as the number of correct links and clusters in the used
reference mappings. Datasets DS1 and DS3 are used to evaluate the quality of
entity clusters generated by the distributed holistic clustering while DS2, DS4
and DS5 are used to analyze the efficiency and scalability of our approach (see
Section \ref{sec:results}). In the following, we briefly describe the datasets
for the two domains before we describe the novel reference dataset for
multi-source clustering in Section~\ref{sec:benchmark}. 

\subsubsection{Geographic Domain} 

\begin{figure*}[tbp]
    \centering
        \includegraphics[width=1.0\textwidth]{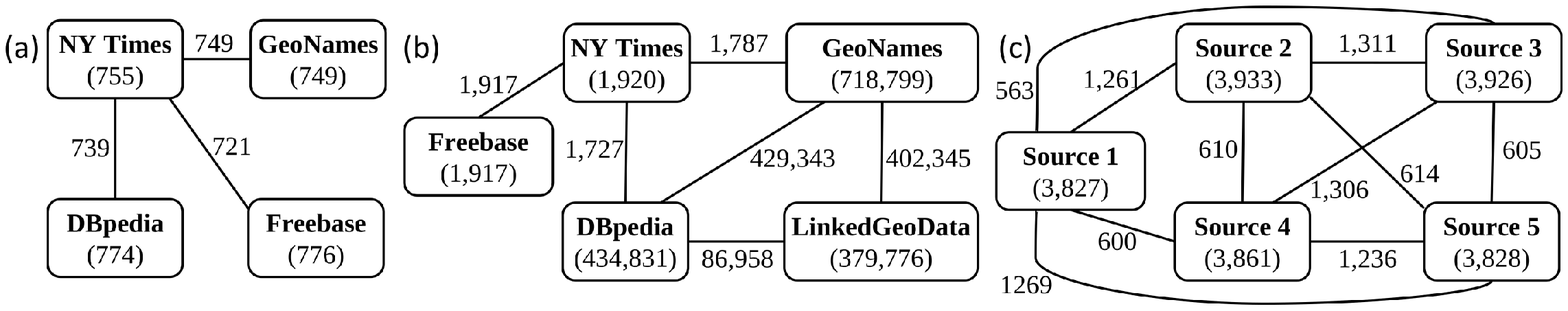}
        \caption{Dataset structures for DS1\,(a), DS2\,(b) and DS3\,(c) with number of entities and links.}
    \label{fig:dataset-sizes}
\vspace{-5mm}
\end{figure*}

We use two datasets (DS1, DS2) from the
geographic domain, covering entities from the data sources DBpedia, GeoNames, NY
Times, Freebase for DS1 and additionally LinkedGeoData for DS2. Entities for
both datasets have been enriched with properties like entity label, semantic
type, and geographic coordinates by using provided SPARQL endpoints or REST APIs.
DS1 is based on a subset of existing links provided by the OAEI 2011 Instance
Matching
Benchmark\footnote{\url{http://oaei.ontologymatching.org/2011/instance/}} that
is also a subset of DS2. \Cref{fig:dataset-sizes} shows the
number of input entities and resulting numbers of links and clusters for the holistic clustering. The $3{,}054$
entities in DS1 create a novel reverence dataset for multi-source clustering
(see \cref{sec:benchmark}). Dataset DS2 covers about 1.5 million
entities from five sources (see \cref{fig:dataset-sizes}\,b) and originates
from the link repository LinkLion~\cite{Nentwig2014}. We reuse about 1 Mio
existing \texttt{owl:sameAs} links from LinkLion as input for the holistic
clustering. 
However, there is no reference dataset available to evaluate the quality of
created clusters for dataset DS2. We use DS2 to evaluate the scalability of our
approach for very large entity sets. 

\subsubsection{Music Domain}

The publicly available Musicbrainz dataset covers artificially adapted entities 
to represent entities from five different data sources~\cite{Hildebrandt2017}\footnote{Musicbrainz test
data \url{https://vsis-www.informatik.uni-hamburg.de/oldServer/teaching//projects/QloUD/DaPo/testdata/}}. 
Every entry in the input dataset represents an audio recording and has
properties like title, artist, album, year, language and length. The property
values have been partially modified and omitted to generate a certain degree of
unclean data and duplicate entities that need to be identified.  This includes
format changes for properties like year (\eg, $'06$, $06$ or $2006$) and song
length (2m 4sec, $02{:}04$, $124000$ or $2.0667$). 

The artificially generated datasets cover between $19{,}375$ and $19{,}375{,}000$ 
tuples (see \cref{tab:datasets}, DS3 to DS5).
Each of the datasets contains a fixed proportion for each cluster size: cluster size
1 ($50\,\%$), size 2 ($25\,\%$), size 3 ($12.5\,\%$), size 4 and 5 ($6.25\,\%$, resp.). 
This means, that for instance $12.5\,\%$ of all entities are in clusters of size 3. 
Beside a set of artificially created duplicates, each dataset covers cluster 
ids from which links between entities, that refer to the same object, 
can be easily derived. Resulting clusters cover between about $16{,}000$ and 
$16{,}000{,}000$ links and up to 10 million clusters. DS3 will be used 
for quality evaluation, while DS4 and DS5 are used to analyze the 
scalability of the distributed holistic clustering. 

\subsection{Reference Dataset for Multi-source Clustering}
\label{sec:benchmark}

\begin{figure}[t!]
\includegraphics[width=\textwidth]{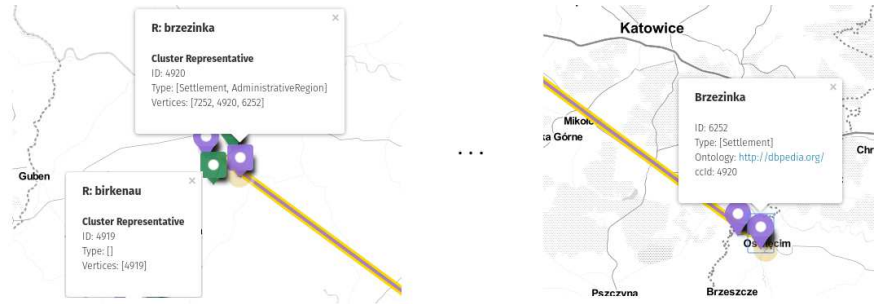}
\caption{Visualization of an incorrect cluster for two different settlements
    named ``Brezinka'' and ``Birkenau'' connected by an incorrect same-as link. Cluster
    representatives are illustrated by rectangles (on the left), while vertices
    are shown as circular pins (on the right).}
\label{fig:birkenau}
\vspace{-4mm}
\end{figure}

We created a new manually curated reference dataset for multi-source clustering 
to support the evaluation of 
holistic clustering approaches w.r.t. the quality of generated clusters. 
Available benchmarks usually only contain links between two data sources. 
 We here provide a gold standard based on real-world data from 
the geographic domain and make it available for other researchers. 
The reference dataset covers the input dataset and the perfect cluster result 
as JSON files and can be downloaded at
\url{https://dbs.uni-leipzig.de/research/projects/linkdiscovery}. 

The reference dataset is a selection of entities 
with the semantic type ``settlement'' from the location subset of the OAEI 2011 Instance
Matching Benchmark.
We made a manual selection decision for vertices using available properties 
and edges. We further checked the correctness of semantic types. For instance, the vertex for
``Canary Island'' was removed since its correct type in the geographic dataset 
should be ``island'' instead of ``settlement''. Removing a vertex resulted in 
the deletion of all its associated edges. 
For manual curation we visualized the data using an
open-source geographic map tool \footnote{uMap \url{https://umap.openstreetmap.fr/en/},
import data as GeoJSON}(see \cref{fig:birkenau}).
To determine the perfect clusters and check for the correctness of
the input data, we used two views on the data. 
First, the original links are represented as (thick) yellow links. 
The second view shows the resulting clusters created by the holistic clustering approach. 
Colored circular pins represent actual vertices
with their properties, while rectangular pins illustrate the newly determined cluster
representatives with label, type, and vertices within the cluster. The
pin color distinguishes different clusters. 

%

In particular, correct clusters were defined based on the results
of the holistic clustering. In case of doubt, Wikipedia was used as additional 
knowledge source about certain geographical entities. 
For each cluster,
we decided whether it was correct or needs to be deleted, 
split into several clusters or merged with another cluster.
During the manual curation, all clusters of maximum size (w.r.t. the number 
of data sources) were correct. Modifications to obtain the perfect result included 
the addition of  further vertices to a cluster or removing vertices from clusters.
For example in ~\cref{fig:birkenau}, 
the original Birkenau/Brzezinka cluster needs to be split into two clusters due 
to a wrong link between two distinct geographic places. In the original
data the cluster covered all four entities. The holistic clustering determined two clusters.
The green pin labeled \emph{Birkenau (Poland)} by NYTimes, as well as the purple pin 
next to it (labeled \emph{Brzezinka} by GeoNames) 
should actually form one cluster. The two vertices of the south-east cluster are another
settlement in Poland which has the same name, \emph{Brzezinka}.
The example represents a difficult case for automatic clustering and linking due to very differing vertex labels. 
The resulting reference dataset covers 820 cluster representatives containing 
information on covered cluster vertices. Based on the cluster ids one can easily 
derive all intra-cluster links to obtain the set of all correct links.

\subsection{Experimental Results}
\label{sec:results}

We now present evaluation results w.r.t. the 
quality of the determined clusters as well as the scalability of the distributed holistic clustering
for the five datasets DS1-DS5. 

\subsubsection{Setup and Configurations}
\label{sec:setup}

The experiments are carried out on a cluster with 16
workers, each of them equipped with a Intel Xeon E5-2430 6x $2.5\,\mathrm{GHz}$,
$48\,\mathrm{GB}$ RAM,
2x $4\,\mathrm{GB}$ SATA disks and $1\,\mathrm{GBit}$ ethernet connection. The machines operate on
OpenSUSE 13.2 using Hadoop 2.6.0 and Flink 1.1.2. All experiments are carried
out three times to determine the average execution time.

In order to obtain input link datasets to apply holistic clustering 
on the geographic dataset DS1, 
we applied LD methods on the input entities using three different 
configurations (config 1, config 2, config 3). 
All configurations compute similarities based on JaroWinkler on the entity
label; configurations 2 and 3 additionally compute a normalized geographic
distance similarity below a maximum distance of $1358\,\mathrm{km}$. Config 1 applies a
minimal similarity threshold of $0.9$ for labels while configs 2 and 3 apply
threshold $0.85$ and $0.9$ for the average label and geographic similarity,
respectively.   
Links for \emph{DS2} were extracted from the LinkLion~\cite{Nentwig2014}
repository, and have been computed by different LD tools from the community.

For the music dataset \emph{DS3} we created input links using a soft
TF/IDF implementation weighted on title ($0.6$), artist ($0.3$) and album
($0.1$) with a threshold of $0.35$.
\emph{DS4} and \emph{DS5} are used to show scalability, therefore, we simply
create edges based on the cluster id from the perfect result by linking 
the first entity of each cluster with all its neighbors.


\subsubsection{Quality}
\label{sec:results:qual}

We now analyze the achieved cluster quality for the geographic dataset DS1 based
on precision, recall and F-measure.  We first use existing input links from the
considered subset in the original OAEI dataset (see
\cref{fig:dataset-sizes}\,a). This manually curated benchmark achieves a
precision of $100\,\%$. However, many links between certain data sources are
missing leading to a reduced recall of only $50\,\%$, and a F-measure of
$66.7\,\%$. With the holistic clustering approach, we achieve very good results
w.r.t. recall ($97.1\,\%$) while preserving a good precision ($99.8\,\%$)
resulting in the F-measure of $98.5\,\%$. This shows that our approach produces
high-quality clusters based on existing input links thereby finding many new
links. 


\begin{table*}[tb]
    \centering
    \setlength{\tabcolsep}{1.4mm}
	\begin{tabular}{cccccccccc}
	    \toprule
	    & \multicolumn{3}{c}{config 1} & \multicolumn{3}{c}{config 2} & \multicolumn{3}{c}{config 3}\\
	    \cmidrule(r){2-4} \cmidrule(r){5-7} \cmidrule(r){8-10}
	    & P                            & R                            & F1                             & P       & R       & F1               & P       & R       & F1 \\
	    \midrule
	    input links                                            & $0.933$                      & $0.806$                      & $\mathbf{0.865}$               & $0.964$ & $0.938$ & $0.951$          & $0.981$ & $0.799$ & $0.881$\\
	    best (star1, star2)                                    & $0.863$                      & $0.844$                      & $0.853$                        & $0.963$ & $0.941$ & $\mathbf{0.952}$ & $0.951$ & $0.838$ & $0.891$\\
	    holistic                                               & $0.903$                      & $0.824$                      & $0.862$                        & $0.913$ & $0.919$ & $0.916$          & $0.968$ & $0.836$ & $\mathbf{0.897}$\\
	    \bottomrule
	\end{tabular}
    \caption{Evaluation of cluster quality for geography dataset DS1 w.r.t. precision (P), recall (R) and F-measure (F1).}
    \label{tab:qualityGeo}
\vspace{-6mm}
\end{table*}

\begin{table*}[t]
    \centering
    \setlength{\tabcolsep}{1.4mm}
    \begin{tabular}{cccc}
        \toprule 
                    & P       & R       & F1\\
        \midrule
        input links & $0.835$ & $0.783$ & $0.808$\\
        holistic    & $0.890$ & $0.861$ & $\mathbf{0.876}$\\
        \bottomrule
    \end{tabular}
    \caption{Evaluation of cluster quality for music dataset DS3 w.r.t. precision (P), recall (R) and F-measure (F1).}
    \label{tab:qualityMusic}
\vspace{-6mm}
\end{table*}

However, as input mappings are not perfect in real-world situations, we used automatically generated
input links for three linking configurations (config 1-3) as described above. 
To evaluate the cluster quality, we further compare our results 
with recently published results from~\cite{Saeedi2017}. 
The work implemented several existing clustering algorithms. Here, we only select the respectively best results 
achieved with two versions of Star clustering~\cite{aslam2004star} (star1, star2). 
It is important to note that in contrast to our holistic clustering approach,
star clustering creates overlapping clusters, thus clusters may
contain duplicates. Besides, star clustering does not create a compact cluster representation.
\cref{tab:qualityGeo} shows results w.r.t. the cluster quality for the computed input links, 
the best result of (star1, star2) and our distributed holistic clustering approach. 
The holistic clustering (F-measure $86.2\,\%$) nearly retains the input link quality ($86.5\,\%$) for
config 1, while best(star1, star2) achieves slightly worst results. 
For config 2, the star2 implementation 
achieves a slightly better F-measure ($95.2\,\%$) compared to the input mapping ($95.1\,\%$). 
For config 3 the holistic clustering improves the quality of the input mapping by 
$1.6\,\%$ w.r.t. F-measure ($89.7\,\%$). 

For the music domain, we evaluate the cluster quality for DS3 using a 
set of computed input links (see setup in \cref{sec:setup}). Overall, the quality of the input links
is lower than for DS1. Due to strongly corrupted entities and more properties, 
DS3 is more difficult to handle. Applying the holistic clustering, we identify a
 quality improvement for both precision and recall, resulting in a
significant increase of F-measure  by approx. $7\,\%$ to $87.6\,\%$ (see in \cref{tab:qualityMusic}) showing 
that our holistic clustering approach is able to handle such unclean data. 

Overall, the holistic approach achieves competitive results although the DS1
dataset  
facilitates achieving relatively good input mappings making it difficult for 
any clustering approach to find additional or incorrect links. 


\subsubsection{Scalability}
\label{sec:results:scal}

\begin{figure*}[tb]
\begin{minipage}{\textwidth}
  	\begin{minipage}{0.45\textwidth}
    	\centering
            \begin{tabular}{ccccc}
                \toprule
                \#workers& pre& dec& merge& total\\
                \midrule
                1         & $312$ & $668$ & $351$ & $1331$\\
                2        & $164$ & $367$ & $268$  & $799$ \\
                4        & $79$ & $231$ & $207$   & $518$ \\
                8        & $45$ & $130$ & $186$   & $361$ \\
                16       & $23$ & $42$ & $162$    & $227$ \\
                \bottomrule
            \end{tabular}
  	\end{minipage}
	\begin{minipage}{0.55\textwidth}
    	\centering    	
    	\includegraphics[width=\linewidth,keepaspectratio]{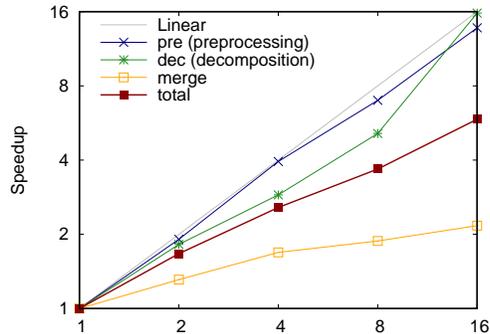}
  	\end{minipage}	
	\caption{Execution times (left) in seconds and speedup (right) for geographic dataset DS2 for the single workflow phases and total workflow.}
        \label{fig:quality-geo}
\end{minipage}
        \vspace{-5mm}
\end{figure*}

To evaluate the distributed holistic clustering w.r.t. efficiency and scalability, 
we determine the absolute execution times as well as the speedup for the very large geographic (DS2) and
music datasets (DS4, DS5). In Flink, several options allow the fine-tuning of parameters
for a distributed cluster environment. In most cases the reasonable way to improve 
 execution times for very large 
datasets is the increase of deployed workers. Another important 
parameter is the parallelism to specify the maximum number of parallel instances of operators or 
data sinks/sources that are available to process data within a Flink workflow. 
We here used the parallelism equal to the number of workers, for which we 
achieved the best execution times. 

\begin{figure*}[tb]
\begin{minipage}{\textwidth}
  	\begin{minipage}{0.45\textwidth}
    	\centering
            \begin{tabular}{ccccc}
                \toprule
                \#workers& pre& dec& merge& total\\
                \midrule
                1        & $423$ &  $419$ &  $608$ & $1450$\\
                2        & $224$ &  $236$ &  $417$ &    $876 $\\
                4        & $121$ &  $123$ &  $301$ &    $545 $\\
                8        & $62$ &   $73$ &   $238$ &    $372 $\\
                16       & $40$ &   $35$ &   $237$ &    $312 $\\
                \bottomrule
            \end{tabular}
  	\end{minipage}
  	\hfill	
	\begin{minipage}{0.55\textwidth}
    	\centering    	
    	\includegraphics[width=\linewidth,keepaspectratio]{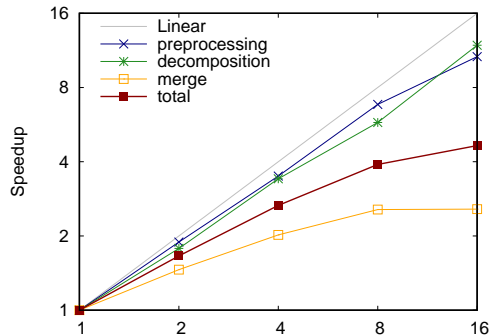}
  	\end{minipage}	
        \vspace{-3mm}
	\caption{Execution times (left) in seconds and speedup (right) for music dataset DS4 for the single workflow phases and total workflow.}
    	\label{fig:quality-music}
\end{minipage}
        \vspace{-5mm}
\end{figure*}

\cref{fig:quality-geo} and \cref{fig:quality-music} show the achieved execution times for DS2 and 
DS4, respectively, for different phases of the clustering workflow as well as the overall workflow execution time. 
For each phase, an increased number of workers leads to improved execution times. For both
domains, the best improvement can be achieved for the
preprocessing (pre) and decomposition (dec) phases. The merge phase is by far more complex.
While preprocessing and decomposition operate within connected
components and clusters, the merge phase attempts to combine similar clusters based
on the assignment in the blocking step and therefore can suffer from data skew problems for some
blocks. These effects become also clear in \cref{fig:quality-geo} and
\cref{fig:quality-music} showing the speedup results compared to the linear optimum. 
Preprocessing and decomposition achieve nearly linear speedup, while the merge phase shows decreased speedup values. 
In total, we achieve a good speedup of $5.86$ for the large 
geographic dataset DS2 and $4.65$ for the large music dataset DS4. 
For the largest dataset DS5 with $\approx$ 20 million entities, we could  determine results for two
configurations: 8 workers could finish the complex task in $43{,}589$ seconds, and 16 workers finished
after $24{,}722$ seconds (reduced by factor $\approx$1.8). 

Overall, the distributed holistic clustering achieves good execution times and moderate scalability results for very large entity sets. The approach is scalable for different data sources and employs a multi-source clustering instead of basic binary linking of two sources. The distributed implementation further allows to scale for a growing number of entities and data sources and is very useful for complex data integration scenarios in big data processing workflows. 

\section{Related Work}
\label{sec:relwork}

Link discovery (LD) has been widely investigated and there are many approaches and
prototypes available as surveyed in \cite{Nentwig2017}. Typical LD approaches
apply binary linking methods for matching two data sources but lack efficient
and effective methods for integrating entities from $k$ different data sources to provide a holistic view for linked data. 
Some approaches enable distributed link discovery or for matching two data sources. For instance, 
Silk MapReduce~\cite{Isele2010} and Limes~\cite{Hillner2011} 
realized LD approaches based on MapReduce before distributed data processing 
frameworks like Spark or Flink became state of the art. 
Similarly, the entity resolution framework Dedoop~\cite{KolbTR12} 
allows to execute complex matching workflows on MapReduce. These tools 
suffer from limitations of MapReduce like repeated data materialization within and between single
jobs and the lack of iterations. They further focus on pairwise matching 
and do not support holistic clustering for multiple data sources or the reuse of existing links sets.

While LD is driven by pairwise linking of data sources, support for multiple
data sources can be found in related research areas.
In~\cite{gruetze2012holistic} ontology concepts from multiple data sources are
 clustered based on topic forests for extracted keywords from concepts and
their descriptions to determine matching concepts within groups of similar topics. 
In~\cite{Megdiche2016} a maximum-weighted graph matching and structural 
similarity computations are applied to concepts of multiple ontologies 
to find high quality alignments. 
However, these holistic ontology matching approaches do not focus on 
clustering of concepts or entities and have limitations w.r.t. 
scalability for LD and many large entity data sources. 

There are few LD approaches that apply clustering or linking for linked data on multiple sources. 
Thalhammer et al.~\cite{Thalhammer2017} present a pipeline for web data fusion 
using multiple data sources applying hierarchical clustering, cluster refinement 
and selection of representatives to achieve a similarity-based clustering with unified entities. 
The unsupervised LD approach Colibri~\cite{Ngomo2014} 
considers error detection and correction for link discovery in multiple knowledge bases 
based on the transitivity of interlinked entities, while clustering of entities is not the main focus. 
These approaches do not realize distributed clustering or linking and have not been evaluated w.r.t. scalability. 
There has been some research for distributed clustering methods 
such as community detection within social networks, \eg, using
MapReduce~\cite{Moon2016} or MapReduce and Spark~\cite{Guo201573}. However these approaches do not focus 
on complex link discovery and data integration workflows, and partially suffer from MapReduce limitations. 
The work in~\cite{Saeedi2017} considers the implementation of 
existing clustering algorithms on top of Apache Flink for entity 
resolution and de-duplication of several data sources. 
The approach does not handle incorrect links and does no use
semantic type information. It further does not create representatives 
for a compact cluster representation as useful for further applications. 

In own previous work~\cite{Nentwig2016}, we proposed a holistic clustering 
approach for multiple Linked Data sources with few initial evaluation results. 
Here we present an extended workflow implementation, in particular the 
realization of holistic clustering in a distributed environment, and 
show comprehensive evaluation results w.r.t. to cluster quality based on a 
novel gold standard, as well as scalability for datasets 
covering very large entity sets from multiple different data sources and domains. 
The created compact cluster representation is particularly useful for 
reuse and incremental cluster extension.



%
%
%







\section{Conclusion}
\label{sec:conclusion}

We presented a distributed holistic clustering workflow for linked data using
the distributed data processing framework Apache Flink. Our approach is based on
the reuse of existing links and is able to handle entities from various
different data sources. We showed the realization of a complex holistic
clustering workflow with dataset transformations and user-defined functions.
Albeit based on Apache Flink, the approach could be also implemented on other
frameworks for distributed data processing such as Apache Spark. We further
provide a novel gold standard for multi-source clustering from the geographic
domain to support the development and evaluation of novel holistic clustering
methods. 
evaluation results for datasets from two different domains with up to 20 million
entities. Our results showed that the proposed approach can achieve a very high
cluster quality. In particular, we were able to find many new correct links and
could remove wrong links within our clustering workflow. We further presented
results for a distributed execution of the holistic clustering in a parallel
cluster environment with very good execution times for the considered dataset
sizes as well as good overall scalability results. 

For future work, we plan to further improve the scalability of our approach,
\eg, by realizing sophisticated blocking and load balancing methods for the
complex cluster merge phase. We further plan the development and combination
with an incremental clustering to support the addition of new entities and data
sources and to particularly address the ongoing growth of the Web of Data and
the accompanied need for incremental holistic clustering methods for LinkedData.

\section*{Acknowledgments}

This research was supported by the Deutsche Forschungsgemeinschaft (DFG) grant
number RA 497/19-2.

\bibliographystyle{splncs03}
\balance
\bibliography{my}
\end{document}